\documentclass[11pt,onecolumn]{article}

\usepackage[utf8]{inputenc}
\usepackage[T1]{fontenc}
\usepackage[a4paper,margin=1in]{geometry}
\usepackage{amsmath,amssymb,amsfonts,amsthm}
\usepackage{mathtools}
\usepackage{bm}
\usepackage{graphicx}
\usepackage{booktabs}
\usepackage{array}
\usepackage{tabularx}
\usepackage{multirow}
\usepackage{longtable}
\usepackage{caption}
\usepackage{subcaption}
\usepackage{xcolor}
\usepackage{enumitem}
\usepackage[hidelinks,colorlinks=true,linkcolor=NavyBlue,citecolor=NavyBlue,urlcolor=NavyBlue]{hyperref}
\usepackage{authblk}
\usepackage{fancyhdr}
\usepackage{float}

\definecolor{NavyBlue}{RGB}{43,58,85}
\definecolor{AccentRed}{RGB}{192,57,43}
\definecolor{AccentGreen}{RGB}{31,138,112}
\definecolor{AccentBlue}{RGB}{46,109,180}
\definecolor{BoxGrey}{RGB}{244,246,249}

\theoremstyle{definition}

\newcommand{\APBTE}{A_{\mathrm{PBTE}}}
\newcommand{\Nstar}{N_{\star}}
\newcommand{\Nref}{N_{\star,\mathrm{ref}}}
\newcommand{\sref}{\sigma_{0,\mathrm{ref}}}
\newcommand{\sigz}{\sigma_{0}}
\newcommand{\ep}{\dot{e}_{p}}
\newcommand{\hd}{\dot{h}_{d}}
\newcommand{\fA}{f_{A}}
\newcommand{\Tsig}{\Theta_{\sigma}}
\newcommand{\dd}{\,\mathrm{d}}
\newcommand{\APBTEi}{A_{\mathrm{PBTE},i}}
\newcommand{\Ahat}{\widehat{A}_{\mathrm{PBTE}}}
\newcommand{\Arate}{\widehat{A}_{\mathrm{rate}}}
\newcommand{\Achrono}{\widehat{A}_{\mathrm{chrono}}}
\newcommand{\fAi}{f_{A,i}}
\newcommand{\sigzi}{\sigma_{0,i}}

\newcommand{\keybox}[1]{%
  \begin{center}\setlength{\fboxsep}{10pt}%
  \colorbox{BoxGrey}{\begin{minipage}{0.92\linewidth}#1\end{minipage}}%
  \end{center}}

\title{A Nonequilibrium Internal-Time Model of Aging: Entropy-Normalized Biological Proper Time and Repair Bifurcations}

\author[ ]{Mesfin Asfaw Taye}
\affil[ ]{West Los Angeles College, Science Division, 9000 Overland Ave, Culver City, CA 90230, USA\\
\texttt{tayem@wlac.edu}}
\date{June 2026 \\[2pt] \small\itshape An application of the Principle of Biological Time Equivalence}

\begin{document}
\maketitle
\thispagestyle{fancy}

% ===================== ABSTRACT =====================
\begin{abstract}
\noindent
Chronological age is an incomplete coordinate for aging. Individuals and species sharing the same calendar time can differ substantially in physiological reserve, molecular damage, mortality hazard, and remaining lifespan. The Principle of Biological Time Equivalence (PBTE) offers a thermodynamic reformulation: biological aging is governed by the accumulation of \emph{internal} physiological time rather than chronological time alone. Building on prior PBTE work, this paper defines the internal-time coordinate $\theta(t)=\int_0^t f(s)\dd s$, where $t$ is chronological time and $f(s)$ is an instantaneous physiological frequency (for example heart rate or respiratory rate), so that $\theta$ is the accumulated count of physiological cycles. Its entropy-normalized extension is $\Tsig(t)=\int_0^t[\sigz(s)/\sref]f(s)\dd s$, where $\sigz(s)=\dd\Sigma/\dd\theta$ is the entropy produced per physiological cycle (the entropy cost per biological tick), $\Sigma$ is cumulative entropy production, and $\sref$ is a fixed reference entropy cost per cycle used as a normalizing unit. The normalized PBTE age $\APBTE(t)=\Tsig(t)/\Nref$ measures the fraction of a reference entropy--cycle budget consumed, where $\Nref$ is the reference number of entropy-weighted cycles available over a lifetime. The manuscript is explicitly theoretical: no empirical cohort is analyzed, and the numerical demonstrations are synthetic stress tests rather than validation. The revised model has three components. First, a linear nonequilibrium damage law, $\dd D/\dd A=\mu+(\lambda-r)D$, is retained as the analytically transparent baseline; here $D$ is an aggregate damage burden, $A\equiv\APBTE$ is PBTE biological age, $\mu>0$ is the baseline damage-production rate per unit PBTE age, $\lambda>0$ is the damage-feedback coefficient, and $r>0$ is the repair rate. Second, saturating repair, $R(D)=rD/(K+D)$, in which $K>0$ is the damage scale at which repair begins to saturate, produces a genuine saddle-node bifurcation: the healthy low-damage fixed point disappears when $r$ falls below $r_c=\mu+\lambda K+2\sqrt{\lambda\mu K}$, giving a dynamical tipping point for runaway senescent damage. Third, the previous algebraic curve-collapse demonstration is replaced by a non-circular numerical reconstruction test: noisy, heterogeneous physiological and entropy-cost proxies are generated, the reconstructed clock $\Ahat$ is rebuilt from those proxies alone, and collapse is tested against independently generated damage and hazard outcomes, with deliberately mis-specified clocks included as failure controls. The resulting framework does not replace molecular theories of aging; it supplies a thermodynamic accounting coordinate, a nonlinear repair-instability mechanism, and a falsifiable empirical protocol for future longitudinal validation.

\end{abstract}

\bigskip
\hrule
\bigskip

% =====================================================================
\section{Introduction}
\label{sec:intro}

Aging is usually indexed by chronological time. A mouse, a dog, an elephant, and a human are all said to age by the same external clock, although the physiological meaning of one year differs radically among them. Even within a single species, individuals of identical chronological age may differ in frailty, disease burden, inflammatory state, epigenetic age, and mortality risk. This mismatch between calendar time and biological state is not a minor measurement problem; it indicates that aging requires an \emph{internal} coordinate.

The standard molecular view of aging has identified many important damage and maintenance processes, including genomic instability, telomere attrition, epigenetic alteration, loss of proteostasis, mitochondrial dysfunction, deregulated nutrient sensing, stem-cell exhaustion, inflammation, and altered intercellular communication~\cite{lopezotin2013,lopezotin2023}. These mechanisms describe \emph{what} changes during aging. They do not by themselves define the clock with respect to which those changes advance. The central proposal of the present paper is that PBTE supplies that missing clock.

In prior PBTE work~\cite{taye_cardio,taye_book,taye_bte_vert,taye_cardiac_invariant,taye_thermo_param,taye_neural}, biological proper time was defined as the accumulated count of physiological cycles. For a recurrent physiological process with frequency $f(t)$, the internal-time coordinate is $\theta(t)=\int_0^t f(s)\dd s$. The same framework showed that a lifetime cycle count can be written as the ratio between total lifetime entropy production and the mean entropy cost of one physiological cycle, $\Nstar=\Sigma/\langle\sigz\rangle$. In the homeostatic regime, $\sigz$ is operationally estimated from metabolic power, body temperature, and physiological frequency as $\sigz\simeq P/(T f)$. This converts the empirical observation of approximate lifetime-cycle invariance into a thermodynamic statement about entropy cost per biological tick. The approximate constancy of lifetime physiological-cycle counts across body sizes is consistent with the long-standing allometric regularities relating metabolic rate, lifespan, and body mass~\cite{west1997,schmidt1984,hulbert2007}, and with comparative longevity and senescence biology~\cite{finch1990,barja1998}.

To keep the notation self-contained, the principal symbols used throughout are defined here. The variable $t$ is chronological (calendar) time and $L$ is the lifespan. The physiological frequency $f(t)$ (equivalently $\fA(t)$) is an instantaneous cycle rate such as heart rate or respiratory rate; its time integral $\theta(t)$ is the accumulated cycle count, i.e.\ biological proper time. The quantity $\ep(t)\ge 0$ is the irreversible entropy-production rate, $\hd(t)\ge 0$ is the entropy exported to the environment, $\Sigma(t)=\int_0^t\ep\dd s$ is cumulative entropy production, and $P(t)$ and $T(t)$ are metabolic power and absolute body temperature. The entropy cost per physiological cycle is $\sigz(t)=\ep(t)/f(t)=\dd\Sigma/\dd\theta$, the thermodynamic price of one biological tick, and $\sref$ is a fixed reference value of $\sigz$ used as a normalizing unit. The total lifetime cycle count is $\Nstar=\theta(L)$, and $\Nref$ is the reference entropy-weighted cycle budget. The entropy-normalized internal time is $\Tsig(t)=\int_0^t(\sigz/\sref)f\dd s$, and PBTE biological age is $\APBTE(t)=\Tsig(t)/\Nref$, abbreviated $A$. In the damage model, $D$ is the aggregate damage burden, $R(D)$ the repair flux, $h$ the mortality hazard, and the parameters $\mu,\lambda,r,K,\alpha,\gamma$ are defined where they first appear. The mortality framing throughout follows the classical Gompertz description of an exponentially rising hazard~\cite{gompertz1825,kirkwood2015}.

This paper applies that framework to aging. The aim is deliberately narrower than a complete theory of senescence. PBTE does not claim that a single scalar variable explains every molecular event in aging. Rather, it proposes that aging processes can be organized by an internal-time coordinate whose rate is set by physiological frequency and entropy production. The claim is not that chronological time is irrelevant, but that chronological time is only the \emph{external} parameter. Biological age is the progress of the organism along its internal thermodynamic trajectory.

The core hypothesis is simple: an organism ages as it spends a finite budget of entropy-normalized biological cycles. A fast physiological pace spends this budget quickly; a slow physiological pace spends it slowly. A high entropy cost per cycle accelerates budget consumption; a low entropy cost per cycle extends the number of cycles that can be sustained before failure. Aging interventions and disease states can therefore be classified by how they alter either the speed of the internal clock or the entropy price of each tick.

\begin{figure}[t]
\centering
\includegraphics[width=\linewidth]{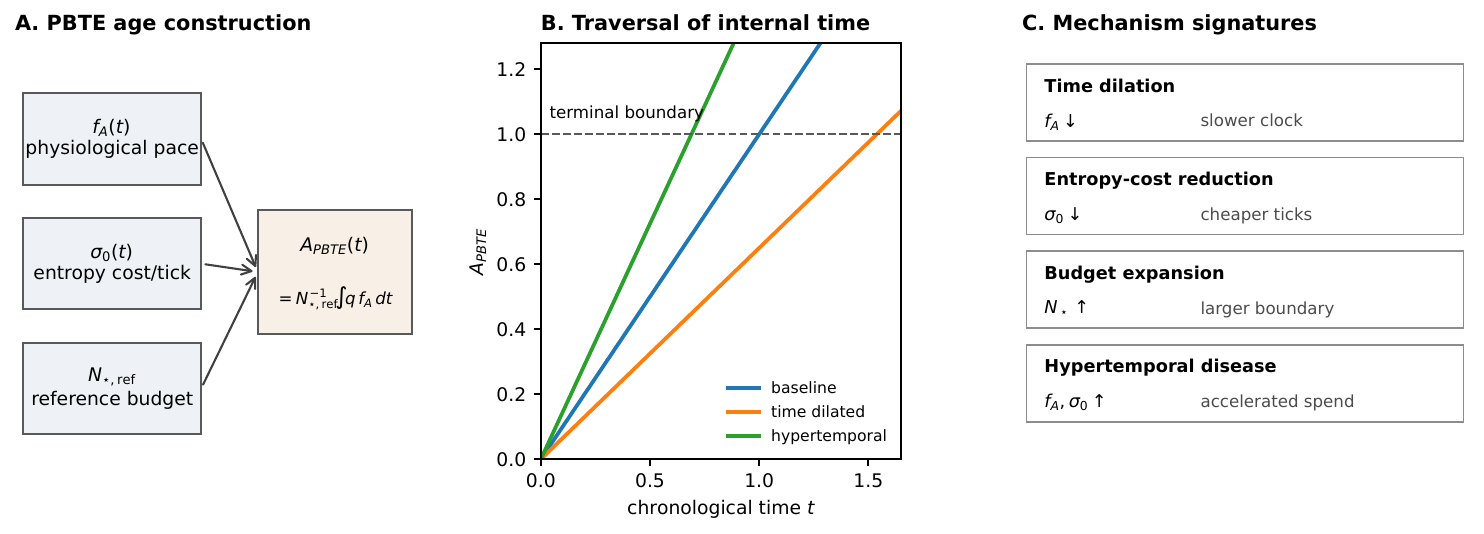}
\caption{\textbf{Schematic of PBTE aging.} \textbf{(A)} Biological age is the entropy-normalized fraction of a finite cycle budget consumed, built from the physiological pace $\fA(t)$, the entropy cost per tick $\sigz(t)$, and the reference budget $\Nref$. \textbf{(B)} Aging is traversal of internal time: a hypertemporal organism reaches the terminal boundary $\APBTE=1$ in fewer calendar years, whereas a time-dilated organism (caloric restriction, torpor) reaches it later. \textbf{(C)} The three mechanism classes defined by their signature in PBTE variables.}
\label{fig:schematic}
\end{figure}

This reformulation has practical consequences. It predicts that biological-age biomarkers, especially DNA methylation clocks, should correlate more tightly with accumulated internal time than with chronological age alone. It predicts that caloric restriction and torpor should slow aging per calendar year primarily by reducing the rate at which biological proper time accumulates~\cite{colman2009,colman2014,mattison2017,carey2003}. It predicts that repair-enhancing or regulation-enhancing mechanisms should reduce aging per physiological cycle. It predicts that chronic inflammation, metabolic syndrome, and neurodegenerative dysfunction should act as hypertemporal states, accelerating biological age by raising frequency, entropy cost, or temporal roughness. These predictions are measurable and falsifiable. Figure~\ref{fig:schematic} summarizes the framework.

% =====================================================================
\section{Previous PBTE Results, Restated and Revised}
\label{sec:relation}

This paper builds on two concrete results from our previous work, which are restated here in revised and self-contained form so that the reader need not consult the earlier papers to follow the present development. We first recall the empirical regularity, then the thermodynamic identity that organizes it, and finally re-derive both from first principles so that every symbol is defined in place.

\paragraph{The elephant and the mouse: why calendar time is the wrong clock.}
The motivating observation is an old one. A mouse lives roughly two to three years, an elephant roughly sixty to seventy, yet the two animals are far more alike when life is measured in heartbeats than in years. A mouse's heart beats several hundred times per minute and an elephant's only about thirty, but each species completes a broadly comparable total number of heartbeats over its natural lifetime. Counted in years the two lives look utterly different; counted in beats they look surprisingly similar. The same holds across warm-blooded vertebrates more generally. This is the empirical seed of the Principle of Biological Time Equivalence (PBTE): the natural clock of an organism is not the external calendar but the internal tally of physiological cycles it executes. A fast small animal and a slow large animal can spend the same internal budget at very different external rates, so that one year of calendar time means something very different to each.

\paragraph{Previous result 1: the empirical lifetime-cycle regularity.}
In our previous work we established that warm-blooded vertebrates show approximate lifetime-cycle regularities~\cite{taye_cardio}. Let $f(t)$ denote an instantaneous physiological frequency (for the cardiac case, the heart rate, in beats per unit time) and let $L$ be the lifespan. The total number of cycles completed over the lifetime is the time integral of the frequency,
\begin{equation}
\Nstar=\int_0^L f(t)\dd t .
\label{eq:Nstar_intro}
\end{equation}
If both the frequency and the lifespan are treated as roughly constant characteristics of a species, this reduces to the elementary estimate $\Nstar\approx f\,L$. For the mouse, a large $f$ and small $L$; for the elephant, a small $f$ and large $L$; yet the product $f\,L$ lands in the same broad range. Empirically, lifetime heartbeat counts concentrate near $\Nstar^{\mathrm{heart}}\sim 10^{9}$ in reference mammals, while respiratory counts are coarser and weaker, of order $\Nstar^{\mathrm{breath}}\sim 2\text{--}3\times 10^{8}$. The revised statement we adopt here is deliberately careful: $\Nstar$ is \emph{not} a strict universal constant but a clade-dependent quantity with a finite spread, and it should always be read as the lifetime integral in Eq.~\eqref{eq:Nstar_intro} rather than as a fixed biological target. The narrowing of $\Nstar$ across mammals is the regularity to be explained, not a law to be assumed.

\paragraph{Previous result 2: the entropy-cost identity, derived from first principles.}
A cycle count alone is descriptive; it does not say \emph{why} the count should be approximately conserved. Our previous work supplied the missing thermodynamic structure~\cite{taye_book}, which we now re-derive so the present paper is self-contained. The organism is treated as an open system that sustains itself by continuously producing and exporting entropy. Let $\ep(t)\ge 0$ be the irreversible entropy-production rate, that is, the entropy generated per unit time by all the irreversible biochemical, transport, mechanical, and regulatory processes that keep the organism alive. The total entropy produced over the lifetime is its time integral,
\begin{equation}
\Sigma=\int_0^L \ep(t)\dd t .
\label{eq:Sigma_intro}
\end{equation}
Biological proper time is the accumulated cycle count $\theta(t)=\int_0^t f(s)\dd s$, so that an infinitesimal advance of internal time is $\dd\theta=f\dd t$. Combining this with the lifetime dissipation of Eq.~\eqref{eq:Sigma_intro}, and dividing the entropy produced in an interval by the number of cycles completed in that same interval, defines the entropy cost of a single cycle,
\begin{equation}
\sigz(t)=\frac{\ep(t)\dd t}{f(t)\dd t}=\frac{\ep(t)}{f(t)}=\frac{\dd\Sigma}{\dd\theta}.
\label{eq:sigma0_intro}
\end{equation}
The quantity $\sigz$ is the thermodynamic price of one biological tick: the entropy an organism must generate to advance its internal clock by one cycle. Writing $\dd\Sigma=\sigz\dd\theta$ and integrating over the whole life gives $\Sigma=\int_0^{\Nstar}\sigz(\theta)\dd\theta$, and dividing by the total cycle count defines the lifetime-average entropy cost per cycle,
\begin{equation}
\langle\sigz\rangle=\frac{1}{\Nstar}\int_0^{\Nstar}\sigz(\theta)\dd\theta=\frac{\Sigma}{\Nstar}.
\label{eq:avg_intro}
\end{equation}
Rearranging Eq.~\eqref{eq:avg_intro} yields the central PBTE accounting identity,
\begin{equation}
\boxed{\;\Nstar=\frac{\Sigma}{\langle\sigz\rangle}\;}.
\label{eq:prev_identity}
\end{equation}
In words: the number of physiological cycles a life contains equals the total entropy produced over that life divided by the average entropy cost of one cycle.

\paragraph{Estimating the entropy cost from measurable quantities.}
Equation~\eqref{eq:sigma0_intro} is exact but abstract, because $\ep$ is not directly measurable. In near-homeostatic conditions it can be estimated from familiar physiological quantities. To leading thermodynamic order, an organism in a metabolic steady state exports entropy to its surroundings at the rate at which it dissipates metabolic heat divided by its absolute temperature. Let $P(t)$ be the metabolic power, that is, the rate at which the organism dissipates free energy as heat (units of energy per time, for example watts), and let $T(t)$ be the absolute body temperature (in kelvin). The entropy exported as heat is then $P/T$, and in steady state this balances the entropy produced, so
\begin{equation}
\ep(t)\simeq\frac{P(t)}{T(t)},
\qquad\text{hence}\qquad
\sigz(t)\simeq\frac{P(t)}{T(t)\,f(t)}.
\label{eq:closure_intro}
\end{equation}
Equation~\eqref{eq:closure_intro} shows that the entropy cost of one physiological cycle is set by metabolic power per unit temperature per unit frequency: a cycle is expensive when the organism burns power fast relative to how often it ticks, and cheap when the same power is spread over many ticks. This closure is what makes $\sigz$, and therefore $\Nstar$, estimable from metabolic and physiological data rather than from entropy directly.

\paragraph{What is exact and what is testable.}
The revised reading separates two claims that the earlier presentation stated together. Equation~\eqref{eq:prev_identity} is an \emph{exact identity}: once cycle count and entropy cost are defined, it follows by construction and carries no empirical risk. The \emph{testable physical claim} is the separate statement that the mass-specific entropy cost per cycle $\langle\sigz\rangle$ is approximately invariant within a defined physiological regime. Given that invariance, the near-constancy of $\Nstar$ follows from Eq.~\eqref{eq:prev_identity} together with allometric cancellation: large animals produce more total entropy $\Sigma$ but also pay a proportionally larger cost per cycle, and the two scalings with body mass largely cancel, leaving $\Nstar$ in a narrow band~\cite{west1997,schmidt1984}. This is the thermodynamic explanation of the elephant--mouse equivalence: the mouse spends a small lifetime entropy budget quickly and the elephant spends a large one slowly, but each spends a comparable number of entropy-normalized cycles. Conflating the definitional identity with the falsifiable hypothesis, as the earlier draft implicitly did, obscures exactly this point; the present paper keeps them distinct.

\paragraph{What the present paper adds.}
The previous work stopped at the lifetime ledger: how many entropy-normalized cycles a life contains. It did not describe how an organism \emph{moves} through that ledger, nor what happens dynamically as the budget is spent. The present paper takes the identity in Eq.~\eqref{eq:prev_identity} as its starting point and asks a new question: if life advances by spending entropy-normalized cycles, how should aging itself be described? Three features carried forward from the earlier results are essential. First, the exact identity is kept distinct from the testable invariance claim, as above. Second, chronological time $t$ is kept distinct from biological proper time $\theta$: $t$ is the external coordinate set by the calendar, while $\theta$ is the accumulated internal cycle count that the organism actually experiences. Third, deviations from the baseline---clade multipliers, disease states, and interventions---are treated as structured physiological effects, interpreted as changes in either the rate of internal-time accumulation, the entropy cost per cycle, or the accessible cycle budget, rather than as noise. The remainder of this paper extends these three ideas into a dynamical theory of aging in which biological age is not time since birth but the fraction of the internal entropy--cycle budget already traversed.

% =====================================================================
\section{Thermodynamic Foundation of PBTE Aging}
\label{sec:thermo}

A living organism is an open nonequilibrium system that maintains
macroscopic order by continuously transforming chemical free energy into
mechanical work, ion gradients, biosynthesis, molecular repair,
signaling, and heat~\cite{schrodinger,prigogine,seifert2012,taye_noneq_stoch,taye_entropy_balance}. It is
therefore not an isolated body that passively degrades in calendar time,
but a regulated dissipative system whose internal state is sustained by
continuous thermodynamic throughput. Let $S(t)$ denote the coarse-grained
internal entropy of the organism, $\ep(t)\ge 0$ the irreversible
entropy-production rate, and $\hd(t)\ge 0$ the entropy exported to the
environment. The open-system entropy balance is
\begin{equation}
\frac{\dd S}{\dd t}
=
\ep(t)-\hd(t).
\label{eq:balance}
\end{equation}
The term $\ep$ represents entropy generated by irreversible biochemical,
transport, mechanical, and regulatory processes, whereas $\hd$ represents
entropy removed through heat dissipation, respiration, circulation,
excretion, and environmental exchange.

In adult homeostasis the organism is not at equilibrium. Rather, it is
maintained near a nonequilibrium steady state in which internal entropy
is regulated within a bounded physiological range. On timescales long
compared with fast metabolic fluctuations but short compared with
lifespan-scale deterioration, Eq.~\eqref{eq:balance} gives
\begin{equation}
\frac{\dd S}{\dd t}\simeq 0,
\qquad
\ep(t)\simeq \hd(t).
\label{eq:homeostatic_balance}
\end{equation}
To leading thermodynamic order, the entropy exported as heat is estimated
by the dissipated metabolic power divided by absolute body temperature,
\begin{equation}
\hd(t)\simeq \frac{P(t)}{T(t)}.
\label{eq:heat_entropy_export}
\end{equation}
Combining Eqs.~\eqref{eq:homeostatic_balance} and
\eqref{eq:heat_entropy_export} gives the homeostatic closure
\begin{equation}
\ep(t)\simeq \hd(t)\simeq \frac{P(t)}{T(t)}.
\label{eq:closure}
\end{equation}
This approximation is not intended to reduce all biological maintenance
to heat flow alone. It is a leading-order thermodynamic closure linking
metabolic throughput, temperature, and entropy production.

Aging becomes visible when this balance is no longer exact. If entropy
export, molecular repair, proteostasis, immune regulation, and cellular
renewal fail to fully compensate irreversible entropy production, the
coarse-grained internal burden associated with disorder, damage, and
loss of regulation accumulates. Integrating Eq.~\eqref{eq:balance} from
the initial life-history time $0$ to age $t$ gives
\begin{equation}
S(t)
=
S(0)
+
\int_0^t
\bigl[\ep(s)-\hd(s)\bigr]\dd s .
\label{eq:Saccum}
\end{equation}
The accumulated entropy in Eq.~\eqref{eq:Saccum} measures the net excess of irreversible production over export. PBTE does not identify this coarse-grained entropy with a single
biomarker of aging. Instead, entropy production is used to assign a
thermodynamic cost to the advancement of an internal physiological clock.

Let $f(t)$ be a physiological frequency, such as heart rate, respiratory
rate, or another biologically meaningful rhythm. The accumulated
biological proper time is defined as the total number of physiological
cycles completed up to age $t$,
\begin{equation}
\theta(t)
=
\int_0^t f(s)\dd s,
\qquad
\dd\theta=f(t)\dd t.
\label{eq:theta}
\end{equation}
Thus $\theta$ is not an external time coordinate; it is an internal
cycle coordinate. If $f$ is the heart rate, $\theta$ is the accumulated
number of heartbeats. If $f$ is the breathing rate, $\theta$ is the
accumulated number of breaths.

The entropy produced in the same infinitesimal interval is
\begin{equation}
\dd\Sigma
=
\ep(t)\dd t .
\label{eq:dSigma_time}
\end{equation}
Converting Eq.~\eqref{eq:dSigma_time} from chronological to internal time and using $\dd t=\dd\theta/f(t)$ from Eq.~\eqref{eq:theta}, this becomes
\begin{equation}
\dd\Sigma
=
\frac{\ep(t)}{f(t)}\,\dd\theta .
\label{eq:dSigma}
\end{equation}
Equation~\eqref{eq:dSigma} identifies the instantaneous entropy cost per biological tick as
\begin{equation}
\sigz(t)
=
\frac{\ep(t)}{f(t)}
=
\frac{\dd\Sigma}{\dd\theta}.
\label{eq:sigma0}
\end{equation}
The quantity $\sigz$ defined in Eq.~\eqref{eq:sigma0} is the entropy produced per physiological cycle. It
is therefore the thermodynamic price of one increment of biological
proper time. A cycle is more expensive when entropy production is high
relative to the physiological frequency, and less expensive when the
same cycle is executed with lower irreversible dissipation.

Let $L$ be the lifespan, defined here as the terminal point of the
life-history interval under consideration. The total entropy production
over the lifespan is
\begin{equation}
\Sigma
=
\int_0^L \ep(t)\dd t,
\label{eq:Sigma_total}
\end{equation}
and the total lifetime cycle count is
\begin{equation}
\Nstar
=
\theta(L)
=
\int_0^L f(t)\dd t .
\label{eq:Nstar_def}
\end{equation}
Equations~\eqref{eq:Sigma_total} and~\eqref{eq:Nstar_def} are the lifetime totals of dissipation and of cycles, respectively. Since $\dd\Sigma=\sigz\,\dd\theta$, the same total entropy production can
be written in the internal-time coordinate as
\begin{equation}
\Sigma
=
\int_0^{\Nstar}\sigz(\theta)\dd\theta .
\label{eq:Sigma_theta}
\end{equation}
Dividing Eq.~\eqref{eq:Sigma_theta} by the cycle count, the lifetime-average entropy cost per biological tick is therefore
\begin{equation}
\langle\sigz\rangle
=
\frac{1}{\Nstar}
\int_0^{\Nstar}\sigz(\theta)\dd\theta
=
\frac{\Sigma}{\Nstar}.
\label{eq:avgsigma}
\end{equation}
Rearranging Eq.~\eqref{eq:avgsigma} gives the PBTE cycle-count relation
\begin{equation}
\boxed{
\Nstar
=
\frac{\Sigma}{\langle\sigz\rangle}
}.
\label{eq:cyclecount}
\end{equation}
Equation~\eqref{eq:cyclecount} is an accounting identity with biological
content. It states that the number of physiological cycles completed
over life equals the total entropy produced divided by the average
entropy cost of one cycle. It should not be interpreted as a claim that
death occurs at an exact universal number of heartbeats, breaths, or
metabolic events. Rather, it states that biological time advances by
consuming an entropy-normalized cycle budget.

The simplest limiting case makes the interpretation transparent. If
$f(t)=f$ and $\sigz(t)=\sigz$ are approximately constant, then
\begin{equation}
\Nstar=fL,
\qquad
\Sigma=\ep L,
\qquad
\sigz=\frac{\ep}{f}.
\label{eq:constant_case}
\end{equation}
Substituting the constant-rate relations of Eq.~\eqref{eq:constant_case} into the cycle-count identity~\eqref{eq:cyclecount} then gives
\begin{equation}
L=\frac{\Nstar}{f}.
\label{eq:L_N_f}
\end{equation}
Equation~\eqref{eq:L_N_f} shows that, for a fixed cycle budget, a higher physiological frequency implies
a shorter chronological lifespan, whereas a lower physiological
frequency implies a longer chronological lifespan. This is the elementary
PBTE intuition behind approximate heartbeat, breath, and metabolic-cycle
invariants.

The metabolic closure also gives a complementary thermodynamic form. Substituting the homeostatic closure of Eq.~\eqref{eq:closure} into the definition~\eqref{eq:sigma0},
\begin{equation}
\sigz(t)
\simeq
\frac{P(t)}{T(t)f(t)}.
\label{eq:sigz_metabolic}
\end{equation}
Equation~\eqref{eq:sigz_metabolic} shows that the entropy cost of one physiological cycle is set by metabolic
power per unit temperature per unit frequency. High metabolic power
increases the thermodynamic cost of each cycle, whereas a higher
physiological frequency distributes the same entropy production over more
cycles.

% =====================================================================
\section{Biological Age as a Normalized Internal-Time Coordinate}
\label{sec:bioage}

The simplest PBTE definition of biological age is the normalized
internal-time coordinate
\begin{equation}
A_{\theta}(t)
=
\frac{\theta(t)}{\Nstar}.
\label{eq:Atheta}
\end{equation}
In Eq.~\eqref{eq:Atheta}, $A_{\theta}=0$ at the beginning of the relevant life-history
interval and $A_{\theta}=1$ at the terminal PBTE boundary. If the
physiological frequency is approximately constant, then
\begin{equation}
A_{\theta}(t)
=
\frac{ft}{\Nstar},
\qquad
L=\frac{\Nstar}{f}.
\label{eq:Atheta_constant}
\end{equation}
In the limit of Eq.~\eqref{eq:Atheta_constant}, chronological age and normalized biological cycle age are
proportional. Organisms with larger physiological frequencies traverse
the same internal cycle distance in fewer calendar years.

Cycle count alone, however, is not sufficient. The same physiological
cycle can have different thermodynamic costs depending on mitochondrial
efficiency, oxidative stress, inflammation, repair capacity, endocrine
state, autonomic stability, temperature, and disease. A heartbeat, breath,
or metabolic cycle executed in a well-regulated organism is therefore
not equivalent to the same cycle executed under pathological stress. To
include this effect, PBTE introduces the entropy-normalized biological
time
\begin{equation}
\Tsig(t)
=
\int_0^t
\frac{\sigz(s)}{\sref}\,
f(s)\dd s ,
\label{eq:Thetasigma}
\end{equation}
where $\sref$ is a reference entropy cost per physiological cycle. A
cycle with $\sigz=\sref$ contributes one reference biological tick. A
cycle with $\sigz> \sref$ contributes more than one effective tick, and a
cycle with $\sigz<\sref$ contributes less than one.

Using the definition $\sigz(t)=\ep(t)/f(t)$, Eq.~\eqref{eq:Thetasigma}
can be rewritten as
\begin{equation}
\Tsig(t)
=
\int_0^t
\frac{\ep(s)}{\sref}\dd s
=
\frac{\Sigma(t)}{\sref},
\label{eq:Thetasigma_entropy}
\end{equation}
where
\begin{equation}
\Sigma(t)
=
\int_0^t \ep(s)\dd s
\label{eq:Sigma_t}
\end{equation}
is the accumulated entropy production up to age $t$. Hence
$\Tsig$, through Eqs.~\eqref{eq:Thetasigma_entropy} and~\eqref{eq:Sigma_t}, has two equivalent interpretations: it is the number of
entropy-weighted physiological cycles, and it is the accumulated entropy
production measured in units of the reference entropy cost per cycle.

The corresponding PBTE biological age is defined by
\begin{equation}
\APBTE(t)
=
\frac{\Tsig(t)}{\Nref},
\label{eq:APBTE}
\end{equation}
where $\Nref$ is the reference entropy-cycle budget. Substituting
Eq.~\eqref{eq:Thetasigma} into Eq.~\eqref{eq:APBTE} gives the central PBTE definition
\begin{equation}
\boxed{
\APBTE(t)
=
\frac{1}{\Nref}
\int_0^t
\frac{\sigz(s)}{\sref}\,
f(s)\dd s
}.
\label{eq:APBTE_central}
\end{equation}
Equivalently, using Eq.~\eqref{eq:Thetasigma_entropy},
\begin{equation}
\APBTE(t)
=
\frac{\Sigma(t)}{\sref\Nref}.
\label{eq:APBTE_entropy}
\end{equation}
Defining the reference lifetime entropy budget
\begin{equation}
\Sigma_{\rm ref}
=
\sref\Nref,
\label{eq:Sigma_ref}
\end{equation}
and inserting Eq.~\eqref{eq:Sigma_ref} into Eq.~\eqref{eq:APBTE_entropy}, one obtains the compact thermodynamic form
\begin{equation}
\boxed{
\APBTE(t)
=
\frac{\Sigma(t)}{\Sigma_{\rm ref}}
}.
\label{eq:APBTE_budget}
\end{equation}
Equation~\eqref{eq:APBTE_budget} states that PBTE biological age is the fraction of a reference lifetime
entropy--cycle budget that has been consumed.

Differentiating Eq.~\eqref{eq:APBTE_central} gives the instantaneous
rate of biological aging,
\begin{equation}
\frac{\dd \APBTE}{\dd t}
=
\frac{1}{\Nref}
\frac{\sigz(t)}{\sref}
f(t).
\label{eq:aging_rate_1}
\end{equation}
Since $\sigz(t)f(t)=\ep(t)$, Eq.~\eqref{eq:aging_rate_1} becomes
\begin{equation}
\frac{\dd \APBTE}{\dd t}
=
\frac{\ep(t)}{\sref\Nref}
=
\frac{\ep(t)}{\Sigma_{\rm ref}}.
\label{eq:aging_rate_2}
\end{equation}
Finally, applying the metabolic closure $\ep(t)\simeq P(t)/T(t)$ to Eq.~\eqref{eq:aging_rate_2},
\begin{equation}
\frac{\dd \APBTE}{\dd t}
\simeq
\frac{P(t)}{T(t)\Sigma_{\rm ref}}.
\label{eq:aging_rate_metabolic}
\end{equation}
Equation~\eqref{eq:aging_rate_metabolic} makes explicit the thermodynamic content of PBTE:
biological aging accelerates when irreversible entropy production
increases and slows when functional maintenance is achieved with lower
entropy production per effective cycle.

\keybox{%
\centering
\textbf{Central definition.}\\[4pt]
$\displaystyle
\APBTE(t)
=
\frac{1}{\Nref}
\int_0^t
\frac{\sigz(s)}{\sref}\,
f(s)\dd s
=
\frac{\Sigma(t)}{\Sigma_{\rm ref}}$\\[4pt]
{\small\itshape biological age $=$ fraction of the reference entropy--cycle budget consumed}}

Equation~\eqref{eq:APBTE_central} shows that an organism ages faster
when the physiological clock runs faster, when each biological tick
becomes more thermodynamically expensive, or when both effects occur
together. It ages more slowly when physiological rate is reduced without
functional collapse, when entropy production per cycle is lowered by
more efficient maintenance, or when the accessible entropy--cycle budget
is extended.

This formulation also explains why chronological age is sometimes
predictive and sometimes misleading. Calendar time is a reliable proxy
for biological state only when $f(t)$, $\sigz(t)$, and the effective
budget are similar across individuals. When these quantities differ, two
organisms of the same chronological age can occupy different positions on
the entropy-normalized internal-time axis.

% =====================================================================
% =====================================================================

% =====================================================================
\section{Minimal Nonequilibrium Aging Model: Linear Baseline}
\label{sec:minimal}

A clock tells time but does not, by itself, explain decline. Having defined the internal-time coordinate, we now ask what accumulates as that clock advances. This section introduces the simplest possible answer---a single aggregate damage variable governed by a linear balance of production, feedback, and repair---and uses it to establish a point that recurs throughout the paper: the speed of the biological clock and the rate of damage along it are logically independent and can be acted upon separately. The linear model is too simple to capture senescence, but its very simplicity isolates this separation cleanly before the nonlinear mechanism of the next section is introduced.

Sections~\ref{sec:thermo} and~\ref{sec:bioage} define $\APBTE$ as an entropy-normalized coordinate of biological proper time. The purpose of the present section is to convert that kinematic definition into a minimal dynamical theory of aging. The model is deliberately coarse-grained: it does not replace molecular accounts of genomic instability, proteostatic failure, mitochondrial dysfunction, inflammation, or cellular senescence. Rather, it provides a common temporal coordinate in which those processes may be represented, compared, and coupled to mortality. Its central methodological advantage is that it separates two conceptually distinct quantities: the rate at which biological time is expended in calendar time and the rate at which damage accumulates per unit biological time.

The state variables are
\begin{equation}
A(t)\equiv \APBTE(t),
\qquad
D(t),
\qquad
h(t),
\label{eq:statevars_revised}
\end{equation}
where $A$ denotes PBTE biological age, $D$ is a nonnegative aggregate damage burden, and $h$ is the instantaneous mortality hazard. The variables in Eq.~\eqref{eq:statevars_revised} are coupled in a definite order: $A$ drives $D$, and $D$ in turn drives $h$. The variable $D$ should be understood as a phenomenological state coordinate summarizing the combined effects of molecular lesions, inflammatory burden, mitochondrial impairment, loss of proteostasis, and regulatory instability. It is not assumed to correspond to a single biomarker. The clock equation follows directly from Eq.~\eqref{eq:APBTE_central}:
\begin{equation}
\frac{\dd A}{\dd t}
=
\frac{1}{\Nref}\frac{\sigz(t)}{\sref}\fA(t)
=
\frac{\ep(t)}{\sref\Nref}.
\label{eq:clock_revised}
\end{equation}
Equation~\eqref{eq:clock_revised} makes chronological time an external parameter rather than the intrinsic coordinate of aging. Calendar time influences the biological trajectory only through the physiological and thermodynamic variables that determine the expenditure rate of internal time. Consequently, chronological damage accumulation may be decomposed through the chain rule as $\dd D/\dd t=(\dd D/\dd A)(\dd A/\dd t)$. This decomposition is important because two organisms may have the same damage kinetics per unit PBTE age but traverse PBTE age at different rates, or conversely may accumulate biological time at the same rate while differing in damage susceptibility per unit internal time.

The baseline damage law is
\begin{equation}
\frac{\dd D}{\dd A}
=
\mu+\lambda D-R(D).
\label{eq:general_damage_revised}
\end{equation}
Here $\mu>0$ denotes baseline damage production per unit PBTE age. The coefficient $\lambda>0$ represents autocatalytic or self-amplifying damage: pre-existing dysfunction can increase inflammatory signaling, metabolic instability, repair demand, proteostatic stress, and regulatory error, thereby raising the subsequent rate of damage production. The repair functional $R(D)$ represents the aggregate action of molecular repair, immune clearance, protein quality control, cellular turnover, tissue remodeling, and compensatory homeostasis. Equation~\eqref{eq:general_damage_revised} therefore expresses aging as a competition among constitutive damage input, damage-dependent positive feedback, and maintenance capacity, all measured relative to the organism's internal thermodynamic clock.

\begin{figure}[H]
\centering
\includegraphics[width=0.76\linewidth]{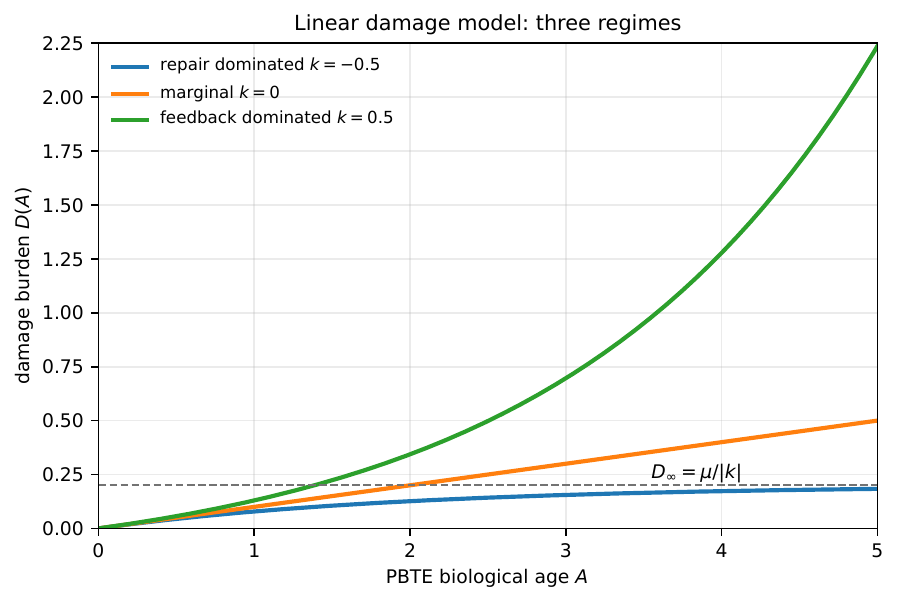}
\caption{\textbf{Linear damage regimes.} Solutions of Eq.~\eqref{eq:linear_damage_revised} with $\mu=0.1$ and $D_0=0$. The repair-dominated regime approaches the bounded value $D_\infty=\mu/|k|$; the marginal regime grows linearly; and the feedback-dominated regime accelerates exponentially with PBTE biological age.}
\label{fig:damage_regimes}
\end{figure}

The simplest analytically tractable closure assumes repair proportional to the current damage burden,
\begin{equation}
R(D)=rD,
\label{eq:linear_repair_revised}
\end{equation}
where $r>0$ is an effective repair coefficient per unit PBTE age. Substituting the linear closure of Eq.~\eqref{eq:linear_repair_revised} into Eq.~\eqref{eq:general_damage_revised} gives
\begin{equation}
\frac{\dd D}{\dd A}=\mu+(\lambda-r)D\equiv \mu+kD,
\qquad
k\equiv \lambda-r.
\label{eq:linear_damage_revised}
\end{equation}
The parameter $k$ is the net damage-amplification rate in internal time. For $k\ne 0$, the solution of Eq.~\eqref{eq:linear_damage_revised} with initial condition $D(0)=D_0$ is
\begin{equation}
D(A)=\left(D_0+\frac{\mu}{k}\right)e^{kA}-\frac{\mu}{k},
\label{eq:linear_solution_revised}
\end{equation}
and for the marginal case $k=0$,
\begin{equation}
D(A)=D_0+\mu A.
\label{eq:linear_marginal_revised}
\end{equation}
Equations~\eqref{eq:linear_solution_revised} and~\eqref{eq:linear_marginal_revised} show that the sign of $k$ partitions the baseline dynamics into three qualitatively distinct regimes:
\begin{align}
k<0\quad(\lambda<r)&:\quad \text{repair dominated, }D\to D_\infty=\mu/|k|,\nonumber\\
k=0\quad(\lambda=r)&:\quad \text{marginal, }D\sim \mu A,\nonumber\\
k>0\quad(\lambda>r)&:\quad \text{feedback dominated, }D\sim e^{kA}.
\label{eq:linear_regimes_revised}
\end{align}
The three cases of Eq.~\eqref{eq:linear_regimes_revised} have distinct biological meaning. In the repair-dominated regime, maintenance compensates for both constitutive and feedback-mediated damage, and the system approaches a bounded nonequilibrium steady state. In the marginal regime, repair exactly balances the linear feedback term, leaving baseline damage production to generate approximately linear deterioration in PBTE age. In the feedback-dominated regime, each increment of existing damage increases the subsequent rate of damage accumulation, producing exponential amplification in internal time. These regimes are shown in Figure~\ref{fig:damage_regimes}.

The value of the linear model is primarily conceptual. It demonstrates that the biological clock and the damage law need not be conflated: Eq.~\eqref{eq:clock_revised} governs how rapidly the organism traverses internal time, whereas Eq.~\eqref{eq:linear_damage_revised} governs how damage evolves along that trajectory. This distinction permits interventions to act either on aging velocity, on damage production and repair per unit internal time, or on both. Nevertheless, the linear closure is biologically restrictive because it assumes that repair can increase without bound as damage rises. Enzymatic repair, immune clearance, proteostasis, stem-cell renewal, and tissue remodeling all operate under finite energetic, molecular, and organizational constraints. The linear boundary $\lambda=r$ therefore provides a useful baseline but not a sufficient theory of senescent loss of resilience. The next section replaces linear repair with a saturating repair law and thereby introduces a genuine threshold and saddle-node bifurcation.

% =====================================================================
\section{Nonlinear Repair and the Senescence Bifurcation}
\label{sec:nonlinear}

Why does aging so often appear gradual for decades and then accelerate abruptly into frailty? The linear model of the previous section cannot reproduce this pattern, because unbounded repair admits no threshold. The essential biological fact it omits is that maintenance has finite capacity: repair enzymes, immune clearance, proteostasis, and tissue renewal all saturate. This section shows that incorporating saturation, through a single change to the repair term, transforms the smooth linear baseline into a system with a genuine tipping point---a stable healthy state that can vanish discontinuously. This is the central dynamical result of the paper, and it gives senescence the character of a loss of resilience rather than a steady wearing-down.

A minimal representation of finite maintenance capacity requires repair to increase approximately linearly at low damage while approaching a finite maximum at high damage. We therefore adopt the saturating closure
\begin{equation}
R(D)=\frac{rD}{K+D},
\qquad r>0,
\qquad K>0.
\label{eq:saturating_repair}
\end{equation}
The parameter $r$ is the maximum repair flux per unit PBTE age, whereas $K$ is the half-saturation damage scale, since $R(K)=r/2$. At fixed $r$ and $D>0$, a larger $K$ implies weaker repair recruitment, while a smaller $K$ implies that repair approaches its maximal capacity at a lower damage burden. Accordingly, $K$ is not intrinsically a beneficial capacity parameter; its physiological interpretation depends on how rapidly repair is mobilized and how the underlying repair function is altered.

\begin{figure}[H]
\centering
\includegraphics[width=\linewidth]{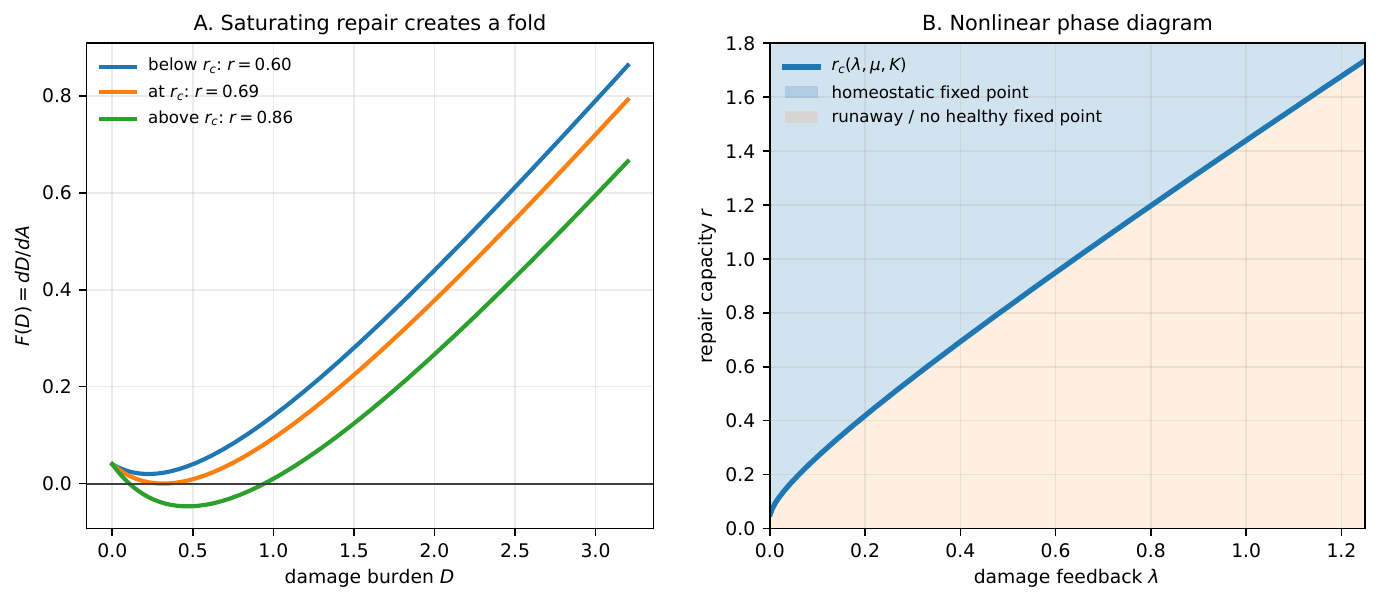}
\caption{\textbf{Nonlinear repair bifurcation.} \textbf{(A)} With saturating repair, $F(D)=\dd D/\dd A$ develops a fold. Above the critical repair capacity $r_c$, a stable low-damage fixed point and an unstable damage threshold coexist. At $r=r_c$ they collide in a saddle-node bifurcation; below $r_c$ no healthy fixed point remains. \textbf{(B)} Nonlinear phase diagram. The curve $r_c=\mu+\lambda K+2\sqrt{\lambda\mu K}$ separates the region in which a low-damage homeostatic state exists from the region of unavoidable damage escalation.}
\label{fig:bifurcation}
\end{figure}

Substituting the saturating repair law of Eq.~\eqref{eq:saturating_repair} into Eq.~\eqref{eq:general_damage_revised} yields
\begin{equation}
\frac{\dd D}{\dd A}
=F(D)
=\mu+\lambda D-\frac{rD}{K+D}.
\label{eq:nonlinear_damage}
\end{equation}
Fixed points of Eq.~\eqref{eq:nonlinear_damage} satisfy $F(D)=0$. Multiplication by $K+D$ gives
\begin{equation}
\lambda D^2+(\mu+\lambda K-r)D+\mu K=0.
\label{eq:fixed_quadratic}
\end{equation}
The discriminant of the quadratic in Eq.~\eqref{eq:fixed_quadratic} is
\begin{equation}
\Delta=(\mu+\lambda K-r)^2-4\lambda\mu K.
\label{eq:discriminant}
\end{equation}
For $\lambda>0$, two positive fixed points require not only $\Delta\ge 0$ in Eq.~\eqref{eq:discriminant} but also sufficiently large repair capacity, so that the sum of the roots is positive. The physically relevant saddle-node condition $\Delta=0$ is
\begin{equation}
(\mu+\lambda K-r)^2=4\lambda\mu K,
\label{eq:saddlenode_condition}
\end{equation}
which, solved for $r$, yields the critical repair capacity defined by the saddle-node condition~\eqref{eq:saddlenode_condition},
\begin{equation}
r_c(\lambda,\mu,K)=\mu+\lambda K+2\sqrt{\lambda\mu K}.
\label{eq:rc}
\end{equation}
When $r>r_c$, the two positive roots of Eq.~\eqref{eq:fixed_quadratic} are
\begin{equation}
D_{\pm}
=
\frac{r-\mu-\lambda K\pm\sqrt{(\mu+\lambda K-r)^2-4\lambda\mu K}}{2\lambda}.
\label{eq:fixed_points}
\end{equation}
The local stability of the branches in Eq.~\eqref{eq:fixed_points} is determined by the sign of
\begin{equation}
F'(D)=\lambda-\frac{rK}{(K+D)^2}.
\label{eq:stability_derivative}
\end{equation}
Evaluating Eq.~\eqref{eq:stability_derivative} on each branch shows that the lower branch $D_-$ is stable and the upper branch $D_+$ is unstable. Initial conditions and perturbations satisfying $D<D_+$ return toward $D_-$, whereas states driven beyond $D_+$ evolve toward runaway damage because the saturating repair flux can no longer compensate constitutive and feedback-mediated production. At $r=r_c$, the stable and unstable fixed points coalesce; for $r<r_c$, no positive homeostatic fixed point remains. Figure~\ref{fig:bifurcation} displays both the fold in $F(D)$ and the resulting phase diagram in the $(\lambda,r)$ plane.

This nonlinear structure is qualitatively stronger than the linear boundary $\lambda=r$. Under linear repair, the balance between feedback and maintenance is global: repair either dominates or fails at every value of $D$. Under saturating repair, maintenance may be fully adequate at low damage yet become insufficient once the burden exceeds a finite threshold. The resulting phase portrait provides a mechanistic representation of senescence as progressive loss of resilience rather than merely gradual damage accumulation. An organism may remain near $D_-$ over an extended interval, recover from modest perturbations, and nevertheless undergo rapid deterioration when declining repair capacity, increasing feedback strength, rising baseline damage production, or an acute external stressor drives the state across $D_+$.

The bifurcation also generates an experimentally relevant early-warning signal. The local recovery time near the stable branch scales approximately as $\tau_{\mathrm{rec}}\sim |F'(D_-)|^{-1}$. As $r\downarrow r_c$, the magnitude of $F'(D_-)$ approaches zero, so recovery from perturbations becomes progressively slower. This critical slowing down predicts that prolonged recovery after infection, exercise, sleep disruption, metabolic challenge, or other standardized perturbations may reveal proximity to the senescence threshold before large changes in baseline damage are evident. In this sense, resilience dynamics may be more informative than static biomarker levels alone.

Equation~\eqref{eq:rc} further distinguishes intervention mechanisms. Increasing the maximum repair flux $r$, reducing the damage-feedback coefficient $\lambda$, or lowering the baseline input $\mu$ moves the system away from the bifurcation boundary. Within the specific Michaelis--Menten form of Eq.~\eqref{eq:saturating_repair}, decreasing $K$ increases repair at any fixed positive damage burden and lowers the critical value $r_c$; more generally, interventions may reshape the repair function in ways not reducible to a single parameter. By contrast, a time-dilation intervention that lowers $\dd A/\dd t$ does not necessarily alter the bifurcation surface in internal-time coordinates, although it delays the chronological time at which the system approaches a given region of the phase portrait. Entropy-cost reduction may act both on the clock equation and, if lower dissipation reduces irreversible molecular burden per unit PBTE age, on the effective parameter $\mu$. The latter coupling is a substantive empirical hypothesis and should not be assumed without measurement.

% =====================================================================
% =====================================================================
\section{Gompertz-like Mortality in PBTE Time}
\label{sec:gompertz}

The damage dynamics of the preceding sections are internal and not directly observed; what demographers measure is mortality. This section builds the bridge between the two, and in doing so recovers a classical law of aging---the Gompertz exponential rise of hazard---as a special case of the PBTE clock. The pay-off is conceptual as well as formal: it shows precisely when chronological age is an adequate proxy for risk and when it must fail, namely whenever individuals expend internal time at different rates. Readers familiar with the Gompertz law will recognize its slope $G$ re-emerging below, but now decomposed into physiological pace, entropy cost, and budget rather than left as an unexplained empirical constant.

The nonlinear damage model developed in the previous section describes how an aggregate damage burden evolves in the internal coordinate of PBTE biological age. That analysis identifies the existence of a low-damage homeostatic branch, an unstable damage threshold, and a saddle-node loss of resilience when repair capacity is no longer sufficient to balance constitutive and feedback-mediated damage production. To connect this dynamical structure to observable aging outcomes, the latent state variable $D(A)$ must be related to measurable endpoints such as mortality, frailty progression, organ failure, or loss of physiological reserve. The present section introduces this connection through a mortality hazard model expressed in PBTE time.

Let $h$ denote the instantaneous mortality hazard. A natural first closure is to assume that risk increases monotonically with accumulated damage. The simplest accelerating form is
\begin{equation}
h(A)=h_0\exp[\alpha D(A)],
\qquad \alpha>0,
\label{eq:hazard_damage}
\end{equation}
where $h_0$ is a baseline hazard and $\alpha$ measures the sensitivity of mortality risk to the aggregate damage coordinate. Equation~\eqref{eq:hazard_damage} should be interpreted as a phenomenological damage-to-risk map rather than a microscopic derivation of mortality. Its role is to express the biologically plausible assumption that a larger burden of molecular lesions, inflammatory dysregulation, mitochondrial impairment, proteostatic failure, cellular senescence, and regulatory instability increases the probability of system-level failure.

This closure is useful because it translates the qualitative regimes of the damage dynamics into qualitative mortality behavior. While the trajectory remains close to the stable low-damage branch $D_-$, the hazard varies slowly because damage is buffered by maintenance and repair. If the trajectory is pushed across the unstable threshold $D_+$, repair saturation prevents recovery, and $D(A)$ rises rapidly. In that regime, Eq.~\eqref{eq:hazard_damage} predicts accelerated mortality risk. Thus, the nonlinear repair model gives a mechanistic interpretation of late-life hazard acceleration: mortality increases sharply not merely because chronological time has passed, but because the organism has approached or crossed a stability boundary in the internal damage dynamics.

It is important, however, to distinguish the damage-mediated closure from a reduced population-level Gompertz model. Equation~\eqref{eq:hazard_damage} uses the explicit intermediate variable $D(A)$:
\begin{equation}
A \longrightarrow D(A) \longrightarrow h(A).
\end{equation}
For survival analysis across populations, it is often useful to adopt a coarser description in which hazard is written directly as an exponential function of PBTE biological age:
\begin{equation}
h(A)=h_0e^{\gamma A},
\qquad \gamma>0.
\label{eq:hazard_internal_revised}
\end{equation}
Here $\gamma$ is the Gompertz-like slope per unit PBTE age. Equation~\eqref{eq:hazard_internal_revised} is not a mathematical consequence of Eq.~\eqref{eq:hazard_damage} in full generality. Rather, it is a reduced phenomenological model that asks whether PBTE age itself can serve as an effective risk coordinate after coarse-graining over tissue-specific damage variables and unobserved heterogeneity.

This distinction is essential. If the damage variable grows approximately linearly in PBTE age over a relevant interval, then Eq.~\eqref{eq:hazard_damage} reduces approximately to an exponential hazard in $A$. If, however, damage grows exponentially in $A$, as in the linear feedback-dominated regime, then Eq.~\eqref{eq:hazard_damage} may generate an exponential-of-an-exponential hazard. Therefore, the direct law $h(A)=h_0e^{\gamma A}$ should be treated as a testable reduced model, not as a universal theorem. Its value lies in providing a compact bridge between PBTE biological time and classical demographic mortality laws.

The connection with classical Gompertz mortality follows by substituting the PBTE clock into Eq.~\eqref{eq:hazard_internal_revised}. The PBTE age is defined by
\begin{equation}
A(t)
=
\frac{1}{\Nref}
\int_0^t
\frac{\sigz(s)}{\sref}
\fA(s)\,\dd s,
\label{eq:PBTE_age_integral_gompertz}
\end{equation}
where $\fA(s)$ is the physiological ticking rate, $\sigz(s)$ is the entropy cost per tick, $\sref$ is the reference entropy cost, and $\Nref$ is the reference entropy-normalized cycle budget. Equivalently, since the entropy-production rate satisfies $\ep(s)=\sigz(s)\fA(s)$, Eq.~\eqref{eq:PBTE_age_integral_gompertz} may be written as
\begin{equation}
A(t)
=
\frac{1}{\sref\Nref}
\int_0^t \ep(s)\,\dd s.
\label{eq:PBTE_age_entropy_integral}
\end{equation}
Equation~\eqref{eq:PBTE_age_entropy_integral} shows that PBTE age is accumulated entropy production measured in units of the reference entropy-cycle budget.

Substitution of Eq.~\eqref{eq:PBTE_age_integral_gompertz} into Eq.~\eqref{eq:hazard_internal_revised} gives
\begin{equation}
h(t)=h_0
\exp\left[
\frac{\gamma}{\Nref}
\int_0^t
\frac{\sigz(s)}{\sref}
\fA(s)\,\dd s
\right].
\label{eq:hazard_calendar_revised}
\end{equation}
Equation~\eqref{eq:hazard_calendar_revised} is the PBTE generalization of the Gompertz law. In the classical Gompertz form, mortality hazard is exponential in chronological time. In the PBTE form, mortality hazard is exponential in accumulated entropy-normalized biological time. Chronological age therefore becomes an imperfect proxy for risk whenever different individuals or species expend internal time at different rates.

The instantaneous logarithmic slope of the hazard in calendar time is obtained by differentiating Eq.~\eqref{eq:hazard_calendar_revised}:
\begin{equation}
\frac{\dd}{\dd t}\ln\frac{h(t)}{h_0}
=
\gamma\frac{\dd A}{\dd t}
=
\gamma
\frac{\sigz(t)\fA(t)}{\sref\Nref}
=
\gamma
\frac{\ep(t)}{\sref\Nref}.
\label{eq:instantaneous_gompertz_slope}
\end{equation}
Equation~\eqref{eq:instantaneous_gompertz_slope} identifies the effective chronological Gompertz slope as a product of the mortality sensitivity $\gamma$ and the velocity of PBTE age. Hence, mortality acceleration in calendar time is predicted to depend not only on elapsed time, but also on physiological pace, entropy cost per cycle, and the accessible entropy-normalized budget.

The ordinary chronological Gompertz law is recovered as a special limiting case. If $\fA(t)=\fA$ and $\sigz(t)=\sigz$ are approximately constant over the relevant adult interval, then
\begin{equation}
A(t)=
\frac{\sigz\fA}{\sref\Nref}t.
\label{eq:A_linear_time_revised}
\end{equation}
Inserting Eq.~\eqref{eq:A_linear_time_revised} into Eq.~\eqref{eq:hazard_internal_revised} yields
\begin{equation}
h(t)=h_0e^{Gt},
\qquad
G=
\gamma
\frac{\sigz\fA}{\sref\Nref}.
\label{eq:G_revised}
\end{equation}
Thus, the classical Gompertz slope $G$ is reinterpreted as the product of a risk-sensitivity coefficient and the PBTE aging velocity. Chronological age is expected to function as an adequate risk coordinate only when the velocity of internal time is nearly constant and sufficiently homogeneous across the population under study. When $\fA(t)$, $\sigz(t)$, or $\Nref$ varies across individuals, the same chronological duration corresponds to different distances traveled in PBTE age, and mortality trajectories should separate in calendar time.

This formulation also provides a direct falsification criterion. If PBTE age is the more appropriate temporal coordinate, then survival curves, frailty trajectories, physiological decline, or damage biomarkers that are heterogeneous in chronological time should show improved alignment when expressed as functions of $A$. Such alignment need not be perfect, because tissue-specific aging, competing risks, genetic heterogeneity, environmental exposures, and measurement error may remain. Nevertheless, a PBTE-based survival model should improve out-of-sample likelihood, calibration, discrimination, or trajectory collapse relative to models based only on chronological age or on unintegrated physiological variables. Failure to obtain such improvement would argue against the reduced hypothesis that PBTE age is an effective population-level mortality coordinate.

The central consequence of Eq.~\eqref{eq:G_revised} is that the apparent Gompertz slope can be altered through several distinct mechanisms. A reduction in mortality acceleration may arise from slower physiological ticking, lower entropy cost per tick, a larger accessible reference budget, or changes in the damage--repair phase portrait. These alternatives can produce similar changes in aging per calendar year while representing different biological processes. The next section uses this observation to classify intervention mechanisms by their PBTE signatures.

% =====================================================================
\section{Intervention Classes and Aging Velocity}
\label{sec:interventions}

Why should two interventions that extend lifespan by the same amount be regarded as biologically different? This section answers that question by showing that the PBTE clock decomposes any change in the rate of aging into a small number of mechanistically distinct factors, each with its own measurable fingerprint. This matters because the language of ``slowed aging'' is otherwise ambiguous: it conflates interventions that slow the physiological clock, interventions that make each tick thermodynamically cheaper, interventions that enlarge the available budget, and pathological states that do the reverse. Making these distinctions explicit converts a vague qualitative claim into a set of quantitative, falsifiable hypotheses.

The PBTE expression for the chronological Gompertz slope,
\begin{equation}
G=
\gamma
\frac{\sigz\fA}{\sref\Nref},
\label{eq:G_intervention_start}
\end{equation}
provides a compact algebra for distinguishing mechanistically different ways of modifying aging. Here $G$ is the per-year mortality-acceleration rate, $\gamma$ is the sensitivity of log-hazard to PBTE age, $\fA$ is the physiological ticking rate, $\sigz$ is the entropy cost per tick, and $\Nref$ is the reference entropy-cycle budget. The essential point is that an observed reduction in biological aging per calendar year is not by itself sufficient to identify the underlying mechanism. The same decrease in the apparent slope $G$ may result from reducing the physiological ticking rate $\fA$, reducing the entropy cost per tick $\sigz$, increasing the accessible budget $\Nref$, changing the mortality sensitivity $\gamma$, or modifying the nonlinear damage parameters $(\mu,\lambda,r,K)$. PBTE therefore separates clock-directed mechanisms from damage-directed mechanisms.

A useful first classification follows from the clock equation
\begin{equation}
\frac{\dd A}{\dd t}
=
\frac{\sigz(t)\fA(t)}{\sref\Nref}.
\label{eq:clock_for_interventions}
\end{equation}
Interventions that alter $\dd A/\dd t$ change the rate at which internal biological time is expended in calendar time. Interventions that alter $\dd D/\dd A$ change the amount of damage produced or repaired per unit PBTE age. These two possibilities are mathematically distinct. An organism may spend internal time more slowly while preserving the same damage kinetics per unit $A$, or it may spend internal time at the same rate while reducing the damage burden accumulated per unit $A$. Distinguishing these cases is crucial for interpreting anti-aging interventions. Table~\ref{tab:mechanism_classes_revised} summarizes the resulting four mechanism classes together with their characteristic per-year and per-tick signatures, which the remainder of this section derives.

\begin{table}[t]
\centering
\caption{Mechanism classes distinguished by their PBTE signatures. The principal discriminator is the relation between biological aging per calendar year and biological aging per physiological tick.}
\label{tab:mechanism_classes_revised}
\small
\renewcommand{\arraystretch}{1.28}
\begin{tabularx}{\linewidth}{@{}p{3.2cm} p{3.0cm} X@{}}
\toprule
\textbf{Class} & \textbf{Primary effect} & \textbf{Candidate examples and predicted signature} \\
\midrule
Time dilation & Reduces $\fA$ & Caloric restriction, torpor, hibernation, or physiologically coherent bradycardia; reduced aging per calendar year with approximately preserved aging per tick. \\
Entropy-cost reduction & Reduces $\sigz$ & Improved mitochondrial coupling, proteostatic efficiency, or reduced chronic inflammatory load; reduced aging per physiological tick. \\
Budget expansion & Increases $\Nref$ & Sustained enhancement of repair coordination, systemic resilience, or regulatory reserve; displacement of the effective terminal boundary. \\
Hypertemporal disease & Raises $\fA$, $\sigz$, or both & Chronic inflammation, metabolic dysregulation, neurodegenerative dysfunction, or proliferative pathology; accelerated expenditure of internal time. \\
\bottomrule
\end{tabularx}
\end{table}

The first class is time dilation. If an intervention reduces the physiological ticking rate according to
\begin{equation}
\fA\to c\fA,
\qquad 0<c<1,
\label{eq:fA_time_dilation_transform}
\end{equation}
while leaving $\sigz$, $\Nref$, and $\gamma$ approximately unchanged, then Eq.~\eqref{eq:fA_time_dilation_transform} substituted into Eq.~\eqref{eq:G_intervention_start} gives
\begin{equation}
G\to cG.
\label{eq:timedilation_revised}
\end{equation}
The interpretation of Eq.~\eqref{eq:timedilation_revised} is that the organism traverses PBTE age more slowly per unit calendar time. In this case, aging per calendar year decreases because fewer physiological ticks are accumulated. However, the damage or entropy cost associated with each tick need not be reduced. The defining empirical signature is therefore a reduction in biological aging per calendar year with approximately preserved biological change per physiological cycle. Candidate examples include torpor, hibernation, some forms of caloric restriction, and coherent reductions in cardiovascular or metabolic pace. This mechanism must be distinguished from pathological slowing, in which physiological rate falls because the system is failing. A genuine time-dilation intervention should reduce internal-time velocity while preserving or improving functional capacity.

The second class is entropy-cost reduction. If an intervention reduces the entropy cost per physiological tick,
\begin{equation}
\sigz\to q\sigz,
\qquad 0<q<1,
\label{eq:sigz_cost_transform}
\end{equation}
while leaving $\fA$, $\Nref$, and $\gamma$ approximately unchanged, then Eq.~\eqref{eq:sigz_cost_transform} substituted into Eq.~\eqref{eq:G_intervention_start} gives
\begin{equation}
G\to qG.
\label{eq:cost_reduction_revised}
\end{equation}
According to Eq.~\eqref{eq:cost_reduction_revised}, the number of physiological cycles per unit calendar time may remain the same, but each cycle is thermodynamically cheaper. The predicted signature is therefore reduced biological aging per physiological tick, not merely reduced aging per calendar year. Candidate mechanisms include improved mitochondrial coupling, lower electron leakage, reduced chronic inflammatory signaling, improved proteostatic efficiency, and more efficient cellular maintenance. These examples remain hypotheses unless $\sigz$ or a defensible entropy-production proxy is estimated independently of $\fA$.

The third class is budget expansion. If the accessible entropy-normalized reference budget increases according to
\begin{equation}
\Nref\to b\Nref,
\qquad b>1,
\label{eq:Nref_budget_transform}
\end{equation}
then Eq.~\eqref{eq:Nref_budget_transform} substituted into Eq.~\eqref{eq:G_intervention_start} gives
\begin{equation}
G\to \frac{G}{b}.
\label{eq:budget_revised}
\end{equation}
The scaling in Eq.~\eqref{eq:budget_revised} is conceptually different from slowing the clock. The organism may spend internal time at the same instantaneous rate, but the effective terminal boundary is displaced outward. Budget expansion would correspond to an increased ability to sustain entropy-normalized physiological activity before terminal loss of viability. Possible biological interpretations include improved systemic resilience, enhanced repair coordination, greater regulatory reserve, or more robust tissue-level compensatory capacity. Because $\Nref$ is not directly observable during an individual's lifetime, this class requires careful non-circular inference from longitudinal survival, resilience, recovery, and reserve measurements.

The fourth class is hypertemporal disease. In such a state, physiological pace, entropy cost, or both are increased. If
\begin{equation}
\fA\to c\fA,
\qquad
\sigz\to q\sigz,
\qquad c>1,\quad q>1,
\label{eq:hyper_transform}
\end{equation}
then Eq.~\eqref{eq:hyper_transform} substituted into Eq.~\eqref{eq:G_intervention_start} gives
\begin{equation}
G\to cqG.
\label{eq:hyper_revised}
\end{equation}
By Eq.~\eqref{eq:hyper_revised}, the system spends internal biological time faster, either because the physiological clock ticks more rapidly, because each tick is more dissipative, or because both occur simultaneously. Chronic inflammation, metabolic syndrome, autonomic dysregulation, neurodegenerative disease, and proliferative pathology are candidate examples. However, classification should be empirical rather than diagnostic. Some disease states may reduce externally observed activity while increasing entropy cost per effective physiological cycle. Thus, the sign of $\fA$ and $\sigz$ cannot be inferred from clinical appearance alone.

A central identifiability problem follows directly from Eq.~\eqref{eq:clock_for_interventions}. PBTE age depends on the product
\begin{equation}
\sigz(t)\fA(t)=\ep(t).
\label{eq:product_identifiability}
\end{equation}
Consequently, Eq.~\eqref{eq:product_identifiability} implies that measurements of $A(t)$, $h(t)$, or the effective slope $G$ alone cannot determine whether a change arose from altered physiological pace or altered entropy cost per tick. Time dilation and entropy-cost reduction are observationally equivalent at the level of the integral unless $\fA(t)$ and $\ep(t)$ are estimated separately. In principle, wearable physiological measurements may inform $\fA(t)$, whereas indirect calorimetry, body temperature, metabolic flux measurements, and state-dependent thermodynamic corrections may inform $\ep(t)$. The entropy cost per tick may then be reconstructed as
\begin{equation}
\sigz(t)=\frac{\ep(t)}{\fA(t)}.
\label{eq:sigz_reconstruction}
\end{equation}
Without the independent measurement that Eq.~\eqref{eq:sigz_reconstruction} requires, one cannot determine whether an apparent reduction in aging velocity reflects fewer ticks, cheaper ticks, or both.

The intervention taxonomy also maps onto the nonlinear damage phase diagram. Clock-directed interventions change the speed at which trajectories are traversed in calendar time, whereas phase-diagram interventions change the geometry of the damage dynamics in internal time. In the nonlinear model,
\begin{equation}
\frac{\dd D}{\dd A}
=
\mu+\lambda D-\frac{rD}{K+D},
\label{eq:nonlinear_intervention_recall}
\end{equation}
repair-enhancing interventions increase $r$, feedback-suppressing interventions reduce $\lambda$, interventions that lower constitutive injury reduce $\mu$, and changes in repair recruitment alter $K$ or, more generally, the functional form of $R(D)$. Within the specific saturating closure $R(D)=rD/(K+D)$, the critical repair capacity is
\begin{equation}
r_c(\lambda,\mu,K)
=
\mu+\lambda K+2\sqrt{\lambda\mu K}.
\label{eq:rc_intervention_recall}
\end{equation}
The phase-portrait reading of Eqs.~\eqref{eq:nonlinear_intervention_recall} and~\eqref{eq:rc_intervention_recall} is direct: reducing $\mu$ or $\lambda$ moves the system away from the senescence bifurcation. Decreasing $K$ increases repair recruitment at fixed positive damage and lowers the required critical repair capacity within this particular functional form. Increasing $r$ moves the system upward in the phase diagram and can restore a low-damage homeostatic branch if the system is not too far beyond the bifurcation boundary.

This distinction clarifies why interventions that appear superficially similar may have different biological meanings. A time-dilation intervention may reduce aging per calendar year without changing the internal damage phase portrait. A repair-enhancing intervention may leave the PBTE clock velocity unchanged while increasing the size of the basin of attraction around the low-damage state. An entropy-cost intervention may act on both levels if lower dissipation reduces the velocity of PBTE age and also lowers irreversible damage production per unit internal time. The latter coupling is a substantive empirical hypothesis and should be tested rather than assumed.

Proliferative and metabolically dysregulated states such as cancer may be investigated as localized hypertemporal processes in which tissue-level growth control, substrate use, and dissipation become partially decoupled from organism-level regulation~\cite{hanahan2011,vanderheiden2009}. This interpretation is a proposed application of PBTE, not an established consequence of the minimal model. More generally, the framework provides a testable classification scheme: interventions should be evaluated by their separate effects on $\fA$, $\sigz$, $\Nref$, $\gamma$, and the nonlinear damage parameters $(\mu,\lambda,r,K)$, rather than being grouped indiscriminately under the single description of ``slowed aging.''

The practical implication is that aging interventions should be analyzed in two complementary coordinate systems. In calendar time, one asks whether the observed trajectory of mortality, frailty, or damage is slowed. In PBTE time, one asks why it is slowed: fewer physiological ticks, lower entropy cost per tick, a larger accessible entropy-cycle budget, reduced mortality sensitivity, or a changed damage--repair stability structure. This separation is the main advantage of the PBTE formulation, because it converts the broad language of slowed aging into a set of mechanistic hypotheses that can be measured, compared, and falsified.

% =====================================================================
\section{Non-Circular Numerical Reconstruction Test}
\label{sec:reconstruction}

The previous version of this manuscript contained a curve collapse in which the plotted hazard was defined directly as $h=h_0e^{\gamma A}$ and then displayed against $A$. That is mathematically correct but not evidential: the collapse follows from the definition of the plotted variables. The appropriate numerical test is stricter. Internal time must be reconstructed from noisy physiological and entropy-cost measurements, and only then compared with outcomes that were not used to build the clock.

\subsection{Synthetic physiology and noisy reconstruction}
\label{subsec:synthetic}

For subject $i$, define the latent physiological pace and entropy-cost multiplier
\begin{equation}
q_i(t)=\frac{\sigzi(t)}{\sref},
\qquad
\APBTEi(t)=\frac{1}{\Nref}\int_0^t q_i(s)\fAi(s)\dd s.
\label{eq:latent_A}
\end{equation}
The latent internal time $\APBTEi$ defined in Eq.~\eqref{eq:latent_A} is the ground truth that the reconstruction must recover. The numerical experiment generates heterogeneous, time-varying $\fAi(t)$ and $q_i(t)$ with subject-level variation, smooth fluctuations, transient deviations, and measurement noise. An experimentalist does not observe the latent variables. Instead, the observed proxies are
\begin{equation}
\widehat f_{A,i}(t_j)=\fAi(t_j)\exp(\epsilon^{f}_{ij}),
\qquad
\widehat q_i(t_j)=q_i(t_j)\exp(\epsilon^{q}_{ij}),
\label{eq:noisy_proxies}
\end{equation}
with independent zero-mean noise. From the noisy proxies of Eq.~\eqref{eq:noisy_proxies}, the reconstructed PBTE age is then computed by quadrature from the proxy variables alone:
\begin{equation}
\widehat{A}_{\mathrm{PBTE},i}(t_m)
=
\frac{1}{\Nref}\sum_{j=1}^{m-1}
\frac{\widehat q_i(t_j)\widehat f_{A,i}(t_j)+\widehat q_i(t_{j+1})\widehat f_{A,i}(t_{j+1})}{2}
\Delta t.
\label{eq:Ahat_quad}
\end{equation}
No mortality hazard, damage marker, methylation age, frailty score, or survival endpoint is used in Eq.~\eqref{eq:Ahat_quad}.

To test whether the reconstruction has content, an independent outcome is generated from the latent internal time,
\begin{equation}
Y_i(t_j)=\log h_i(t_j)=a_i+\gamma\APBTEi(t_j)+\eta_{ij},
\label{eq:synthetic_outcome}
\end{equation}
where $a_i$ is a subject effect and $\eta_{ij}$ is independent outcome noise. Crucially, the outcome in Eq.~\eqref{eq:synthetic_outcome} is built from the latent $\APBTEi$ of Eq.~\eqref{eq:latent_A}, not from the reconstructed $\widehat A$ of Eq.~\eqref{eq:Ahat_quad}. This still does not validate the theory biologically; it only asks whether noisy physiological reconstruction can recover the correct latent ordering under controlled conditions.

\subsection{Wrong-clock controls}
\label{subsec:wrongclocks}

A useful numerical demonstration must include failure modes. Two deliberately mis-specified clocks are used:
\begin{equation}
\widehat{A}_{\mathrm{chrono},i}(t)=\frac{t}{T_{\mathrm{ref}}},
\qquad
\widehat{A}_{\mathrm{rate},i}(t)=\frac{1}{\Nref}\int_0^t \widehat f_{A,i}(s)\dd s.
\label{eq:wrong_clocks}
\end{equation}
The first control clock in Eq.~\eqref{eq:wrong_clocks} assumes that calendar time is the aging coordinate. The second includes physiological rate but omits entropy cost. If PBTE is doing nontrivial work, the reconstructed clock in Eq.~\eqref{eq:Ahat_quad} should align outcomes better than both controls when heterogeneity in $q_i(t)$ matters.

\begin{figure}[H]
\centering
\includegraphics[width=\linewidth]{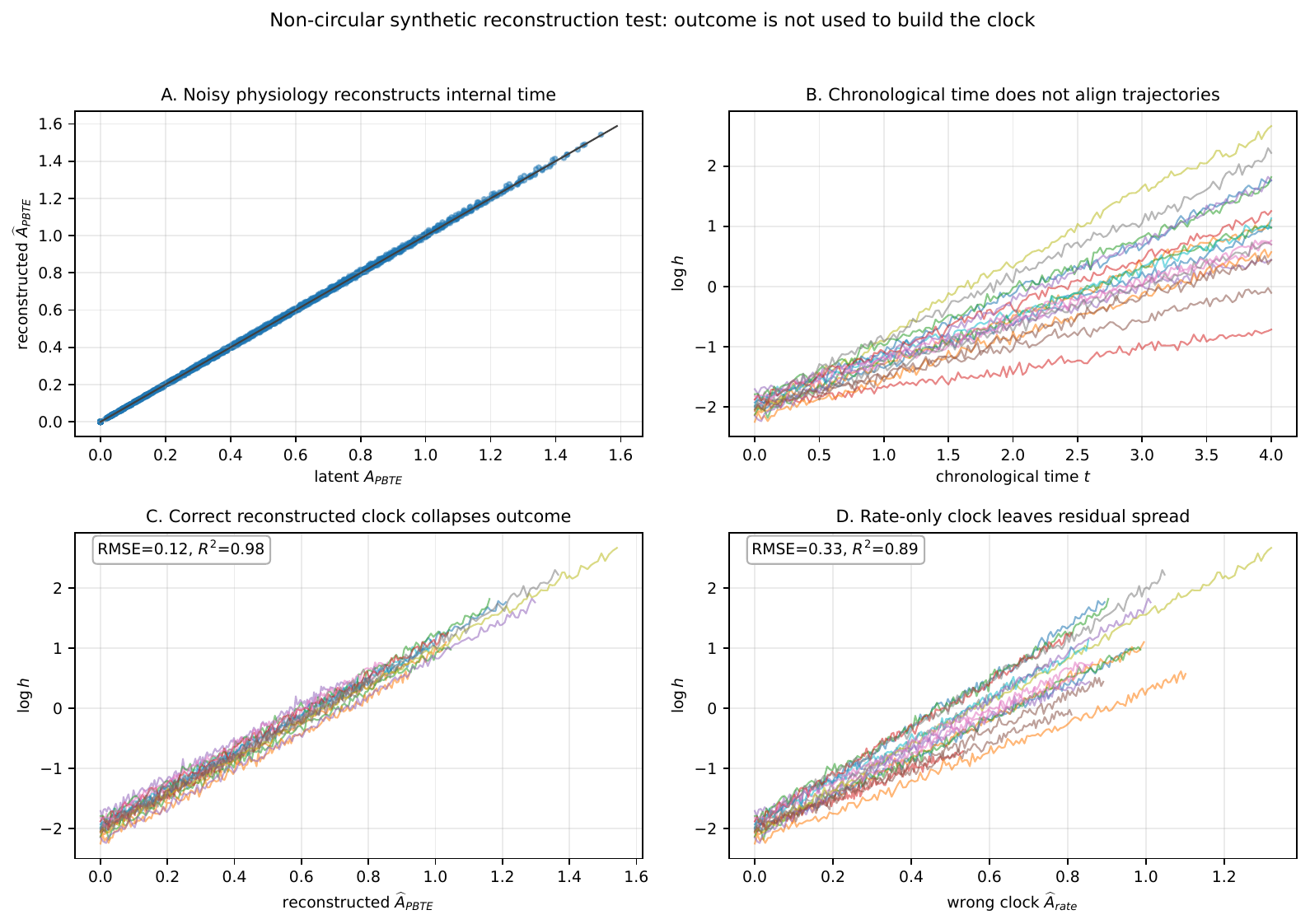}
\caption{\textbf{Non-circular synthetic reconstruction test.} Heterogeneous latent physiology generates $\fA(t)$ and $q(t)=\sigz(t)/\sref$; noisy proxies reconstruct $\Ahat$ without using the outcome. \textbf{(A)} Reconstructed PBTE age tracks the latent internal-time coordinate despite measurement noise. \textbf{(B)} The independently generated log-hazard outcome is dispersed in chronological time. \textbf{(C)} Plotting against reconstructed PBTE age collapses the trajectories. \textbf{(D)} A rate-only wrong clock, which omits entropy cost, leaves substantially larger residual spread. For the deterministic synthetic run shown here, residual RMSE for $\log h$ is $0.124$ using reconstructed PBTE age, $0.453$ using chronological age, and $0.329$ using the rate-only clock. These numbers are illustrative model diagnostics, not empirical validation.}
\label{fig:reconstruction}
\end{figure}

Collapse quality can be quantified by fitting the same model against different candidate clocks,
\begin{equation}
Y_i(t_j)=a+bZ_i(t_j)+\varepsilon_{ij},
\label{eq:collapse_regression}
\end{equation}
where in Eq.~\eqref{eq:collapse_regression} the candidate coordinate $Z$ is $\Ahat$, $\Achrono$, or $\Arate$, and then comparing held-out error, likelihood, calibration, or residual variance. Equivalently, one can bin by a candidate coordinate $Z$ and measure the remaining between-subject spread,
\begin{equation}
Q(Z)=\frac{1}{M}\sum_{m=1}^{M}\operatorname{Var}_i\!\left[Y_i\,\middle|\,Z_i\in B_m\right].
\label{eq:collapse_quality}
\end{equation}
A successful internal-time coordinate should reduce the residual spread $Q(Z)$ of Eq.~\eqref{eq:collapse_quality} relative to chronological and wrong-clock controls. Figure~\ref{fig:reconstruction} reports the outcome of this procedure for a representative synthetic run, contrasting the reconstructed clock with the chronological and rate-only controls.

The central point is methodological. A collapse is meaningful only when the clock is built without using the outcome that is later claimed to collapse. In real data, $\Ahat$ must be pre-specified from wearable physiology, indirect calorimetry, temperature, respiratory variables, inflammatory burden, and other entropy-cost proxies, then tested on independent methylation, frailty, morbidity, and survival endpoints.

% =====================================================================
\section{Testable Predictions and Empirical Validation Strategy}
\label{sec:validation}

The present paper does not claim empirical validation. It defines what validation would require. The decisive test is not whether PBTE can be made to fit data after the fact, but whether a pre-specified, physiology-derived internal-time coordinate improves prediction of independent aging outcomes out of sample.

For each subject, one constructs
\begin{equation}
\widehat{A}_{\mathrm{PBTE},i}(t)=\frac{1}{\Nref}\int_0^t \widehat q_i(s)\widehat f_{A,i}(s)\dd s,
\qquad
\widehat q_i=\widehat\sigma_{0,i}/\sref.
\label{eq:empirical_Ahat}
\end{equation}
The pace variable $\widehat f_{A,i}$ in Eq.~\eqref{eq:empirical_Ahat} can be estimated from wearable heart rate, respiratory rate, activity, temperature, circadian regularity, and HRV~\cite{li2017}. The entropy-cost multiplier $\widehat q_i$ can be approximated using indirect calorimetry, body temperature, inflammatory markers, mitochondrial or metabolic biomarkers, disease state, and possibly methylation or proteomic features if they are not also the endpoint being predicted. The validation rule is strict: variables used to define $\Ahat$ cannot also be counted as independent outcomes.

A survival specification is
\begin{equation}
h_i(t)=h_0(t)\exp\!\left[\beta_A\widehat{A}_{\mathrm{PBTE},i}(t)+\beta_X X_i\right],
\label{eq:survival_revised}
\end{equation}
where $X_i$ includes standard covariates. In the proportional-hazards form of Eq.~\eqref{eq:survival_revised}, PBTE is supported only if $\beta_A>0$ and if models using $\Ahat$ improve pre-specified metrics relative to chronological age, rate-only clocks, cost-only clocks, and established biological-age predictors. Relevant metrics include out-of-sample likelihood, C-index, calibration, prediction error, and collapse quality. For methylation and frailty outcomes, the analogous test is whether $\Ahat$ improves prediction beyond chronological age and covariates without reusing the same biomarkers on both sides of the model. Table~\ref{tab:predictions_revised} collects five concrete falsifiable predictions, each paired with the data required to test it and the specific result that would refute it.

\begin{table}[t]
\centering
\caption{Falsifiable predictions and required validation design for the revised PBTE aging model.}
\label{tab:predictions_revised}
\small
\renewcommand{\arraystretch}{1.32}
\begin{tabularx}{\linewidth}{@{}p{0.8cm} X X X@{}}
\toprule
\textbf{\#} & \textbf{Prediction} & \textbf{Required data} & \textbf{Falsification condition} \\
\midrule
P1 & Physiology-derived $\Ahat$ collapses damage, frailty, methylation, or hazard trajectories better than chronological age. & Longitudinal wearables, entropy-cost proxies, and independent aging outcomes. & Cross-validated collapse is no better than chronological age. \\
P2 & A clock that includes both rate and entropy cost outperforms rate-only and cost-only clocks. & Wearable physiology plus calorimetry, temperature, inflammatory, or metabolic markers. & Omitting $q(t)$ does not degrade prediction when cost heterogeneity is present. \\
P3 & Intervention classes separate by per-year versus per-tick signatures. & Caloric restriction, torpor, exercise, disease, or drug-intervention cohorts with repeated physiology and biomarkers. & Time dilation and entropy-cost reduction produce indistinguishable per-tick effects. \\
P4 & Systems approaching the nonlinear repair threshold show accelerated damage accumulation and increased sensitivity to perturbation. & Longitudinal damage/repair markers, inflammatory markers, frailty, morbidity transitions. & Damage remains linear and reversible with no threshold-like behavior near predicted repair saturation. \\
P5 & Survival prediction improves when $\Ahat$ is pre-specified before outcome testing. & Cohorts with survival follow-up and independent physiology-derived clocks. & $\Ahat$ fails to improve likelihood, calibration, or discrimination after covariate adjustment. \\
\bottomrule
\end{tabularx}
\end{table}

A cross-species version can use pan-mammalian methylation clocks~\cite{lu2023} to test whether species alignment improves when age is expressed in PBTE internal time rather than calendar time. A within-species version can use longitudinal human cohorts with wearables and methylation clocks~\cite{horvath2013,levine2018,lu2019}. In both cases, the validation must be prospective or cross-validated. Estimating $\Nref$ from lifespan and then claiming to predict lifespan would be circular and is not a valid test.

% =====================================================================
\section{Summary and Conclusion}
\label{sec:summary}

This paper has developed a thermodynamic theory of aging in which biological age is identified not with elapsed calendar time but with accumulated, entropy-normalized biological proper time. The central object of the theory is the internal clock
\begin{equation}
\APBTE(t)=\frac{1}{\Nref}\int_0^t\frac{\sigz(s)}{\sref}\fA(s)\dd s
=\frac{\Sigma(t)}{\Sigma_{\mathrm{ref}}},
\label{eq:conclusion_clock}
\end{equation}
which expresses the principal conceptual shift of the framework: from age as \emph{duration} to age as \emph{expenditure}. Chronological age records how long an organism has existed; the coordinate in Eq.~\eqref{eq:conclusion_clock} records what fraction of a finite reference entropy--cycle budget it has consumed. This distinction is not merely semantic. It explains why a single intervention may carry entirely different biological meanings depending on its thermodynamic signature. Slowing physiological rate can extend calendar lifespan when function is preserved, yet a slow organism that pays a high entropy cost per tick need not age slowly; conversely, a metabolically active organism need not age rapidly if efficient maintenance keeps the entropy cost per cycle low and holds the repair system away from its instability boundary. The framework thereby supplies the missing clock with respect to which molecular mechanisms of aging---genomic instability, proteostatic failure, mitochondrial dysfunction, inflammation, and cellular senescence---advance, without claiming to replace those mechanisms or to reduce them to a single scalar.

The dynamical content of the theory rests on three results, developed in Sections~\ref{sec:minimal}--\ref{sec:reconstruction} and summarized here. First, the linear damage--feedback law
\begin{equation}
\frac{\dd D}{\dd A}=\mu+(\lambda-r)D
\label{eq:conclusion_linear}
\end{equation}
is retained as an analytically transparent baseline. Equation~\eqref{eq:conclusion_linear} cleanly separates the rate at which internal time is expended in calendar time from the rate at which damage accumulates per unit internal time, and it organizes the dynamics into repair-dominated, marginal, and feedback-dominated regimes. Its limitation is structural: with repair assumed proportional to damage, the model possesses only the linear phase boundary $\lambda=r$ and cannot represent the finite capacity of real maintenance systems.

Second, replacing proportional repair with the saturating closure $R(D)=rD/(K+D)$ produces a genuine senescence bifurcation,
\begin{equation}
\frac{\dd D}{\dd A}=\mu+\lambda D-\frac{rD}{K+D},
\qquad
r_c=\mu+\lambda K+2\sqrt{\lambda\mu K}.
\label{eq:conclusion_bifurcation}
\end{equation}
For $r>r_c$ in Eq.~\eqref{eq:conclusion_bifurcation}, a stable low-damage homeostatic state coexists with an unstable threshold; at $r=r_c$ the two collide in a saddle-node bifurcation, and for $r\le r_c$ no healthy fixed point survives. This fold geometry furnishes a dynamical-systems interpretation of senescence as a loss of resilience rather than mere gradual decay: an organism may remain near the low-damage branch and recover from modest perturbations for an extended interval, then deteriorate rapidly once declining repair capacity, rising feedback, increasing baseline injury, or an acute stressor drives the state across the threshold. The associated critical slowing down---the divergence of recovery time as $r\downarrow r_c$---offers an experimentally accessible early-warning signal that may precede large changes in static biomarker levels. The model is deliberately generic and identifies no single molecular threshold; its value lies precisely in showing how many distinct mechanisms may share the same fold.

Third, the paper replaces the tautological curve collapse of the earlier draft, in which a hazard defined as $h=h_0e^{\gamma A}$ was plotted against $A$, with a non-circular reconstruction protocol. A clock must be rebuilt from noisy physiological and entropy-cost proxies and only then tested against outcomes that were not used in its construction. The synthetic demonstration of Section~\ref{sec:reconstruction} serves strictly as a proof of procedure: it shows that reconstruction from noisy proxies can recover the latent internal-time ordering, and that deliberately mis-specified chronological and rate-only clocks fail by comparison. It is not, and is not claimed to be, biological validation.

Taken together, these elements situate PBTE within, rather than against, the broader landscape of aging biology. The view of aging as motion along an intrinsic, non-chronological time axis is consistent with classical treatments of the geometry of biological time~\cite{winfree1980} and with the role of circadian organization in healthy aging~\cite{panda2021}, and it is naturally compatible with energy-budget and free-energy accounts of physiological regulation~\cite{friston2010,herculano2011}, in which the maintenance of low entropy production is itself a regulated objective. At the molecular scale, the same stochastic-thermodynamic constraints that govern Brownian motors, fluctuation-driven engines, and the trade-offs among current, activity, and entropy production~\cite{taye_brownian_motors,taye_hybrid_motor,taye_universal_ineq,taye_dopant} fix the entropy price $\sigz$ of each physiological cycle, connecting the cellular engines that sustain homeostasis to the organism-level clock of Eq.~\eqref{eq:conclusion_clock}. The framework is likewise compatible with the diversity of empirical biological-age clocks: developmental methylation clocks, mortality-associated clocks, proteomic and inflammatory clocks, and frailty indices need not measure a single quantity, and PBTE predicts that they may instead be different projections of one internal thermodynamic trajectory, some more sensitive to physiological pace, others to entropy cost, and others to the damage--feedback state.

The reach of these conclusions is bounded by limitations that should be stated plainly, since they define the agenda for subsequent work. The manuscript is a theoretical contribution with numerical illustrations and supplies no empirical validation; the reconstruction experiment is synthetic and demonstrates a testing procedure rather than a biological result. The physiological frequency $\fA(t)$ is not uniquely defined, because aging engages cardiac, respiratory, metabolic, circadian, endocrine, immune, and tissue-specific rhythms whose effective aggregate must be inferred. Entropy production is difficult to measure \emph{in vivo}: the closure $\ep\simeq P/T$ is adequate near homeostasis but incomplete during fever, inflammation, mitochondrial uncoupling, proliferation, and other non-steady states, so the multiplier $q(t)=\sigz(t)/\sref$ will in practice be a proxy carrying uncertainty. The budgets $\Nstar$ and $\Nref$ are effective, class- or individual-specific quantities with finite spread, not rigid universal constants. The saturating repair law is minimal rather than unique, and cooperative, delayed, or tissue-coupled alternatives may yield related but distinct bifurcation structures. Validation must scrupulously avoid circularity: lifespan cannot serve both to calibrate the budget and to claim predictive success, and biomarkers used to construct $\Ahat$ cannot simultaneously be treated as independent endpoints. Finally, the framework is a research instrument and not a clinical recommendation, since lowering physiological rate may be beneficial, neutral, or harmful depending on function, repair, entropy cost, and disease context. These constraints narrow the claim, but in doing so they sharpen it.

In conclusion, PBTE does not displace molecular theories of aging. It contributes a thermodynamic internal-time coordinate on which those processes can be timed, a nonlinear repair mechanism by which senescence acquires the character of a tipping transition, and a concrete, falsifiable protocol by which the entire construction can be tested against longitudinal physiological, molecular, frailty, and survival data. Whether a pre-specified, physiology-derived PBTE clock outperforms chronological age and existing biological-age predictors out of sample is an empirical question that the framework is designed to make answerable, and that remains open.

% =====================================================================
\section*{Declarations}

\paragraph{Data availability.} No new empirical dataset is analyzed in this theoretical manuscript. The figures are generated from the analytic and synthetic models defined in the text. The reconstruction experiment uses simulated noisy physiological and entropy-cost proxies and is intended only as a reproducible numerical stress test.

\paragraph{Competing interests.} The author declares no competing interests.

\paragraph{Funding.} No external funding is declared for this draft manuscript.

\paragraph{Author contributions.} M.A.T. conceived the PBTE aging framework. The revised manuscript develops the theoretical formulation, nonlinear repair bifurcation, numerical reconstruction protocol, illustrative simulations, and validation strategy.

\paragraph{Ethics statement.} No new human or animal subjects research is performed in this theoretical manuscript.

% =====================================================================

\end{document}